\documentclass[12pt,a4paper]{article}
\usepackage[margin=3cm]{geometry}
\usepackage{graphicx}
\usepackage{color}
\usepackage{natbib}
\usepackage{amsmath}
\usepackage{adjustbox}
\usepackage{eurosym}
\usepackage{url}
\usepackage{fancyhdr}

\newcommand{\final}[1]{{#1}}
\usepackage{graphicx}
\usepackage{natbib}
\usepackage{bm}
\usepackage{amsmath}
\usepackage{rotating}
\usepackage{amssymb}
\usepackage{enumerate}

\usepackage{adjustbox}

\newcommand{\E}{\mathbb{E}}
\newcommand{\V}{\mathbb{V}}
\let\proglang=\textsf
\newcommand{\pkg}[1]{{\fontseries{b}\selectfont #1}}
\newcommand\code{\bgroup\@makeother\_\@makeother\~\@makeother\$\@codex}
\def\@codex#1{{\normalfont\ttfamily\hyphenchar\font=-1 #1}\egroup}

\usepackage{amsthm}
\usepackage{amsmath}
\usepackage{natbib}
\usepackage[colorlinks,citecolor=blue,urlcolor=blue,filecolor=blue,backref=page]{hyperref}
\usepackage{graphicx}

\usepackage{amsmath}
\usepackage{amssymb}
\usepackage{algorithm}
\usepackage{booktabs}
\usepackage{subcaption}
\usepackage{color}

\usepackage{multibib}
\newcites{App}{References}

\numberwithin{equation}{section}
\theoremstyle{plain}

\include{newcommands}

\begin{document}
\title{\Large{\centering  How many data clusters are in the Galaxy data set? \\  Bayesian cluster analysis in action}}

\author{Bettina
	Gr\"un\thanks{WU Vienna University of
    Economics and Business}, Gertraud Malsiner-Walli\thanks{WU
    Vienna University of Economics and Business}~{} and \\ Sylvia Fr\"uhwirth-Schnatter\thanks{WU Vienna University of 
    Economics and Business}\\\vspace{0.5cm}}
	
	\date{} 
	
	\maketitle

\sloppy
        
\vspace{-0.5cm}
\begin{small}
  In model-based clustering, the Galaxy data set is often used as a
benchmark data set to study the performance of different modeling
approaches.  \citet{Aitkin:2001} compares maximum likelihood and
Bayesian analyses of the Galaxy data set and expresses reservations
about the Bayesian approach due to the fact that the prior
assumptions imposed remain rather obscure while playing a major role
in the results obtained and conclusions drawn.

The aim of the paper is to address Aitkin's concerns about the
Bayesian approach by shedding light on how the specified priors
influence the number of estimated clusters. We perform a sensitivity
analysis of different prior specifications for the mixtures of
finite mixture model, i.e., the mixture model where a prior on the
number of components is included.  We use an extensive set of
different prior specifications in a full factorial design and assess
their impact on the estimated number of clusters for the Galaxy data
set. Results highlight the interaction effects of the prior
specifications and provide insights into which prior specifications
are recommended to obtain a sparse clustering solution. A simulation
study with artificial data provides further empirical evidence to
support the recommendations.

A clear understanding of the impact of the prior specifications
removes restraints preventing the use of Bayesian methods due to the
complexity of selecting suitable priors. Also, the regularizing
properties of the priors may be intentionally exploited to obtain a
suitable clustering solution meeting prior expectations and needs of
the application. 

\end{small}

\section{Introduction}\label{sec:introduction}

This paper investigates the impact of different prior specifications
on the results obtained in Bayesian cluster analysis based on mixture
models. Mixture models may be used to either approximate arbitrary
densities in a semi-parametric way or in a model-based clustering
context to identify groups in the data. We will focus on the later
application where each component is assumed to potentially represent a
data cluster and the cluster distribution is not approximated by
several mixture components.

\cite{Hennig+Liao:2013} claim that ``there are no unique `true' or
`best' clusters in a data set'' but that the prototypical shape of a
cluster needs to be specified before this question can be answered.
For clustering methods using mixture models, the prototypical shape of
a cluster is in general specified by selecting the component-specific
distributions.  For the fitted mixture model, then a one-to-one
relationship between components and clusters is assumed. For example,
in the case of multivariate metric data one can specify isotropic
Gaussian distributions as component distributions, where the variance
is comparable across components, or Gaussian distributions with
arbitrary variance-covariance matrices, which are allowed to
considerably vary across components \citep[see, for
example,][]{Fraley+Raftery:2002}.

The Bayesian framework provides a principled approach to specify the
prototypical shape of the clusters. By specifying priors on the model
parameters, both the mean prototypical shape as well as the
variability around this prototypical shape are included, i.e., what
the shape on average is as well as how much the component
distributions  vary  across
components. In this sense
the Bayesian approach provides, compared to other clustering methods,
more flexibility to incorporate the prototypical shape of a cluster in
the analysis and hence to  arrive at a suitable clustering solution for
the specific analysis undertaken. In addition the Bayesian framework
also allows to specify a prior on the component weights, thus
influencing if the clusters are a-priori assumed to be rather balanced
in size or if the clustering solution includes both very small and very
large clusters. 
By contrast, for example, $k$-means
clustering assumes that the clusters have an isotropic shape with
similar cluster size and volume \citep[see, for
example,][]{Gruen:2019}.

However, the additional flexibility provided by the Bayesian approach
might also be perceived as overwhelming, in particular, if the
influence of different prior specifications on results obtained
remains rather opaque. \citet{Aitkin:2001} compares maximum likelihood
and Bayesian analyses of mixture models and expresses reservations
about the Bayesian approach due to the fact that the prior assumptions
imposed remain rather obscure while playing a major role in the
results obtained and conclusions drawn. Having sufficient insight into
the influence of prior specifications on the clustering results is
crucial to leverage the advantages of the Bayesian approach where the
priors may be used to regularize the problem and also guide the
analysis to focus on the clustering solution of interest.

In the following we consider the mixture of finite mixture model
(MFM), a name coined by \citet{Miller+Harrison:2018} following
\citet{Richardson+Green:1997}. The MFM is a hierarchical finite
mixture model where a prior on the number of components $K$ is
included. We focus on the MFM, because the Bayesian analysis of the
MFM results in an a-posteriori distribution of the number of data
clusters $K_+$ as well as an a-posteriori distribution of partitions
$\mathcal{C}$. These are both core components of a Bayesian cluster
analysis to address the questions how many data clusters there are in
the data set and how the observations should be grouped into these
data clusters.

Note that in our analyses of the MFM, we make a crucial distinction
between $K$, the number of components in the mixture distribution, and
$K_+$, the number of \textit{filled} components, to which observations
are actually assigned. Only a filled component corresponds to a data
cluster. This implies that, when estimating the number of clusters in
the data, the posterior of $K_+$ is of interest, rather than the
posterior of $K$.  Previously, \citet{Nobile:2004} already
differentiated between $K$ and $K_+$ when analyzing finite mixture
distributions. Also \citet{McCullagh+Yang:2008} made the distinction
between clusters in the population ($K$) and clusters in the observed
sample ($K_+$) and noted that usually a data set contains little
information about the clusters in the population, while being more
informative regarding the number of clusters in the data set.  We will
thus not only investigate the prior on $K$, but also explicitly
inspect the prior on $K_+$, which is induced by the prior on $K$ and
the prior on the mixture weights. In the analysis of the results focus
is given to the posterior of $K_+$ (rather than $K$), determining in
particular the mode of this distribution and its entropy.

We illustrate the impact of different prior specifications using a MFM
of univariate Gaussian distributions for the (in-)famous Galaxy data
set originally introduced to the statistical literature by
\citet{Roeder:1990}. Several results obtained for this data set using
either maximum likelihood estimation or Bayesian analysis methods were
compared and discussed in \citet{Aitkin:2001}.  \citet{Aitkin:2001}
concluded that the maximum likelihood analysis, while having
complications of its own, would be rather straightforward to implement
and be well understood. By contrast, \citet{Aitkin:2001} formulated a
call for action with respect to the Bayesian analysis, asking for a
careful analysis of the role of the priors. This paper aims at
responding to this call for action. Results for the Galaxy data set
are complemented with results of a simulation study with artificial
data to provide further empirical evidence to arrive at
recommendations for suitable prior specifications to obtain a
meaningful clustering result.

\section{Model specification}\label{sec:model-spec-estim}

In our specification of the MFM model with Gaussian component
distributions, the following data generation process is assumed for a
univariate data set of size $n$ given by $\bm{y} = (y_1,\ldots, y_n)$
\citep[see also][]{Richardson+Green:1997}.  One assumes that the
number of components $K$ of the mixture model is sampled from the
prior $p(K)$. Given $K$ the component weights
$\bm{\eta} = (\eta_1, \ldots,\eta_K)$ are sampled from a symmetric
$K$-dimensional Dirichlet distribution with parameter $\gamma_K$. For
each observation $i$ component assignments $S_i$ are drawn from a
multinomial distribution with parameter $\bm{\eta}$.

Regarding the Gaussian component distributions, the component means
$\mu_k$ and the component variances $\sigma^2_k$, $k=1,\ldots,K$, are
independently drawn from the same prior distributions to have
exchangeability.  The component means $\mu_k$ are drawn from a normal
distribution with mean $b_0$ and variance $B_0$, while the component
precisions $\sigma^{-2}_k$, i.e., the inverse variances, are assumed
to follow a Gamma distribution with parameters $c_0$ and $C_0$ (and
expectation $c_0/C_0$). Note that for the prior for the component
distributions not the conjugate prior for the normal distribution with
unknown mean and variance is used, but the independence prior is
employed. If instead the conjugate prior had been used, the
component-specific variances would influence the prior variability of
the component means. This would imply that components which have less
variability also have their mean closer to the prior mean $b_0$. This
prior implication does in general not seem to be appealing in the
mixture context and hence the independence prior is used. For a
further detailed discussion of the priors for the component
distributions see \citet[][Chapter~6]{Fruehwirth-Schnatter:2006}.

Summarizing, this specification results in the following Bayesian
hierarchical MFM model:
\begin{align} \nonumber
K &\sim p(K),\\ \nonumber
\bm{\eta}|K &\sim   \mathcal{D}_K(\gamma_K),\\   \label{eq:MFM}
S_i|\bm{\eta} &\sim \mathcal{M}(\bm{\eta}), \quad  i=1,\ldots,n,\\ \nonumber
\mu_k|b_0,B_0 &\sim \mathcal{N}(b_0,B_0), \quad  k=1,\ldots K,\\ \nonumber
\sigma^{-2}_k|c_0,C_0 &\sim \mathcal{G}(c_0,C_0), \quad  k=1,\ldots K,\\ \nonumber
y_i|\bm{\mu},\bm{\sigma}^2,S_i = k &\sim \mathcal{N}(\mu_k,\sigma^2_k), \quad  i=1,\ldots,n,
\end{align}
where $\bm{\mu} = (\mu_k)_{k = 1,\ldots,K}$ and
$\bm{\sigma}^2 = (\sigma_k^2)_{k = 1,\ldots,K}$.

Additionally, hyperpriors might be specified. For example,
\citet{Richardson+Green:1997} suggest to specify a hyperprior on $C_0$
and \citet{Malsiner-Walli+Fruehwirth-Schnatter+Gruen:2016} add an
additional layer for the prior on the component means which
corresponds to a shrinkage prior allowing for variable selection. In
the following we do not consider adding hyperpriors in order to be
able to assess the influence of different specifications of these
priors and their parameters on the clustering results. In this paper
we focus on the specification of the following priors and parameters:
\begin{itemize}
	\item the prior  $p(K)$ of the number of components $K$,
	\item the value $\gamma_K$ used for the Dirichlet prior,
	\item the prior parameters $b_0$ and $B_0$ for the component means,
	\item the prior parameters $c_0$ and $C_0$ for the component
	variances.
\end{itemize}

\section{The Galaxy data set in statistics}\label{sec:galaxy-data-stat}

The Galaxy data set was originally published in astronomy by
\citet{Postman+Huchra+Geller:1986} and consists of univariate
measurements representing velocities of galaxies, moving away from our
galaxy. In this original publication 83 observations are
listed. \citet{Roeder:1990} introduced the data set to the statistics
literature, but omitted the smallest observation such that in the
following in the statistics literature only 82 observations were
considered. Unfortunately \citet{Roeder:1990} also introduced a typo,
i.e., one observation has a different value than in Table~1 in
\citet{Postman+Huchra+Geller:1986}. A further influential statistics
publication using the Galaxy data set was
\citet{Richardson+Green:1997} who also considered only 82
observations, but corrected the typo and scaled the units by 1000.

The data set was used in statistics by a number of authors to
demonstrate density estimation methods and investigate mixture
modeling approaches.  They either used the version presented by
\citet{Roeder:1990} or by \citet{Richardson+Green:1997}.
%
%
A number of textbooks on applied statistics also use the data set to
demonstrate different statistical methods \citep[see,
e.g.,][]{Lunn+Jackson+Best:2012, Hothorn+Everitt:2014}.

In the following we will use the Galaxy data set as used by
\citet{Richardson+Green:1997}. 
This version of the data set was also used by \citet{Aitkin:2001} when
comparing maximum likelihood and Bayesian analysis methods for
estimating mixture models, focusing in particular on the question of
the number of data clusters in the data set. Within the maximum
likelihood framework, \citet{Aitkin:2001} considered mixtures of
univariate Gaussian distributions with equal as well as unequal
variances. The mixture models were fitted using the EM algorithm
\citep{Dempster+Laird+Rubin:1977} and for each class of component
distributions, the number of components were selected based on the
results of a bootstrap likelihood ratio test
\citep{Aitkin+Anderson+Hinde:1981, McLachlan:1987}. This maximum
likelihood analysis may easily be replicated using the \proglang{R}
package \pkg{mclust} \citep{Scrucca+Fop+Murphy:2016} using
also the Bayesian information criterion (BIC) instead of the
likelihood ratio test for model selection. Based on the maximum
likelihood results, \citet{Aitkin:2001} concludes that ``there is
convincing evidence of three equal variance components, or four
unequal variance components, but no convincing evidence of more than
these numbers, in the velocity data'' (p.~296).

In addition, \citet{Aitkin:2001} reviews the Bayesian analysis of the
Galaxy data set presented in \citet{Escobar+West:1995},
\citet{Carlin+Chib:1995}, \citet{Phillips+Smith:1996},
\citet{Roeder+Wasserman:1997} and
\citet{Richardson+Green:1997}. Table~3 in \citet{Aitkin:2001},
according to its caption, summarizes the posterior distributions of
$K$. However, in fact for the Dirichlet process mixture fitted by
\citet{Escobar+West:1995}, the posterior distribution of $K_+$ is
given. The Bayesian approaches compared differ considerably with
respect to the prior on $K$ and the prior on the component-specific
variances and lead to rather diverse results. \citet{Aitkin:2001}
concludes that some of the Bayesian analysis result in overwhelming
posterior evidence for three groups, while other posterior
distributions obtained are either relatively diffuse over 4--9 with a
mode around 6--7 or are concentrated on the range 7--9. Overall the
cluster solutions for the Galaxy data set are interpreted as either
being sparse, with up to four clusters, or contain many, i.e., more
than four, clusters.

\section{Prior specifications}\label{sec:overv-prior-spec}

In this section, we discuss possible specifications and previous
suggestions in the literature for each of the prior distributions and
their parameters, taking in particular those into account considered
in the Bayesian analysis reviewed in \citet{Aitkin:2001}.  We also
discuss our expectation regarding the effect of these prior
specifications on the cluster solutions obtained, focusing in
particular on the estimated number of data clusters.

\subsection{Prior on $K$}

\citet{Fruehwirth-Schnatter+Malsiner-Walli+Gruen:2020} provide an
overview on previously used priors on $K$ including the uniform
distribution \citep{Richardson+Green:1997}, the truncated Poisson
distribution \citep{Phillips+Smith:1996, Nobile:2004} and the shifted
geometric distribution \citep{Miller+Harrison:2018}. They also propose
the shifted beta-negative-binomial (BNB) distribution as a suitable
alternative which represents a generalization of the Poisson and the
geometric distribution.

Based on this overview, we consider the following priors on $K$:
\begin{itemize}
	\item the uniform distribution $K \sim \text{U}(1, 30)$ with
	prior mean $\E[K] = 15.5$ and prior variance $\V[K] = 74.9$
	\citep{Richardson+Green:1997},
	\item the truncated Poisson distribution
	$K \sim \text{trPois}(3)$ with prior mean $\E[K] = 3.2$
	and prior variance $\V[K] = 2.7$ \citep{Phillips+Smith:1996},
	\item the shifted geometric distribution $K-1\sim \text{Geom}(0.1)$
	with prior mean $\E[K] = 10$ and prior variance $\V[K] = 90$
	\citep{Miller+Harrison:2018},
	\item the shifted BNB distribution $K-1\sim \text{BNB}(1, 4, 3)$ with prior mean
	$\E[K] = 2$ and prior variance $\V[K] = 4$
	\citep{Fruehwirth-Schnatter+Malsiner-Walli+Gruen:2020}.
\end{itemize}

These priors essentially cover all Bayesian MFM analysis reviewed and
compared by \citet{Aitkin:2001}. The only exceptions are
\citet{Carlin+Chib:1995} who perform model selection to decide between
a 3- and a 4-component solution and \citet{Roeder+Wasserman:1997} who
use a uniform distribution with support $\{1, 2, \ldots,
10\}$. \citet{Richardson+Green:1997} point out that the upper bound of
30 for the uniform distribution is inconsequential for their
applications, including the Galaxy data set, because this bound is
never hit during sampling from the posterior distribution. We thus
also use this uniform prior for the Galaxy data set.

The proposed priors for $K$ differ considerably in the prior means and
variances induced. The shifted $\text{BNB}(1, 4, 3)$ has the smallest prior
mean; the truncated Poisson distribution has the smallest prior
variance, with only a slightly higher prior mean. We expect the two
prior distributions $\text{trPois}(3)$ and the shifted $\text{BNB}(1, 4, 3)$,
which have comparable, small means, to induce cluster solutions with
less data clusters compared to the other two priors.  We expect this
behavior to be most pronounced for the truncated Poisson distribution,
because of its lowest variance, thus putting only very little mass on
large values of $K$, e.g., the probability of $K > 10$ is less than
0.001.

\subsection{Prior parameter $\gamma_K$ for the component weights}

All Bayesian MFM analyses considered in \citet{Aitkin:2001} are based
on a MFM with $\gamma_K \equiv 1$.  However, as will be demonstrated
in Section~\ref{sec:Kp}, the Dirichlet parameter $\gamma_K$ crucially
affects the prior on $K_+$, since it determines how closely the prior
on $K_+$ follows the prior on $K$. A more detailed discussion on the
specification of $\gamma_K$ for the MFM is given in
\citet{Fruehwirth-Schnatter+Malsiner-Walli+Gruen:2020}.

\citet{Fruehwirth-Schnatter+Malsiner-Walli+Gruen:2020} suggest to use
an arbitrary sequence for the Dirichlet parameter $\gamma_K$ which
might depend on the number of components $K$. They distinguish two
special cases: the static MFM where $\gamma_K \equiv \gamma$ and the
dynamic MFM where $\gamma_K = \alpha / K$. \citet{McCullagh+Yang:2008}
already discussed these two special cases indicating that they are
structurally different. While previous applications of the MFM focused
on the static case, the Dirichlet process mixture model is included in
the dynamic case.

In the following we will consider the static as well as the dynamic
MFM, with $\gamma \in \{0.01, 1, 10\}$ in the static case and
$\alpha \in \{0.01, 1, 10\}$ in the dynamic case.
Thus, in addition to the popular choice $\gamma \equiv 1$, we consider
also a much smaller value of $\gamma$ and $\alpha$ as well as a much
larger value. The much smaller value is expected to induce a sparse
cluster solution with only very few data clusters and thus also
achieve a certain independence of the specification of the prior on
$K$. The much larger value is expected to induce cluster solutions
with rather equally sized data clusters and also a stronger link
between the number of data clusters and the number of components,
which implies a larger influence of the prior on $K$ in this
setting. We expect that the dynamic MFM leads to sparser solutions
than the static MFM given that $\gamma_K =\alpha/K$ is likely to
assume small values for large $K$.

\subsection{Induced prior of the number of data clusters $K_+$}\label{sec:Kp}

As the posterior of $K_+$, the number of filled components, is the aim
of the analysis, it is illuminating to study the prior on $K_+$. The
prior on $K_+$ is implicitly induced through the specification of the
prior on $K$ and the prior parameter $\gamma_K$.
\citet{Fruehwirth-Schnatter+Malsiner-Walli+Gruen:2020} and
\citet{Greve+Gruen+Malsiner-Walli:2020} present formulas to derive
this implicit prior in a computational efficient way. We investigate
the prior on $K_+$ induced by the prior specifications on $K$ and
$\gamma_K$ considered for the Galaxy data set to further gauge our
prior expectations of the influence of these prior specifications on
the cluster solutions obtained.

\begin{figure}
	\centering
	\includegraphics[width=0.94\textwidth]{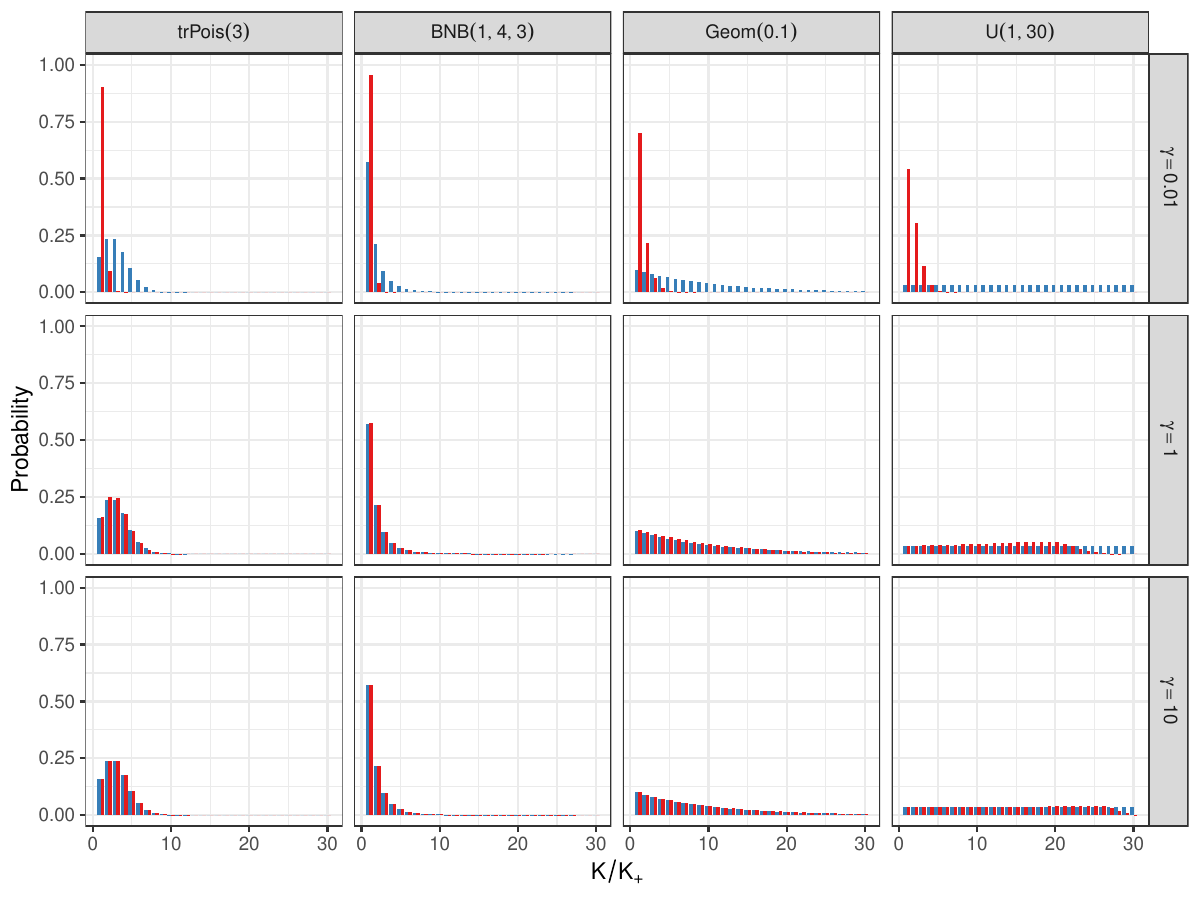}
	\caption{\label{fig:priorKplusStatic} The prior probabilities of $K$
		(in blue) and $K_+$ (in red) for the static MFM for different
		priors on $K$ and values for $\gamma$ with $n = 82$.}
	\includegraphics[width=0.94\textwidth]{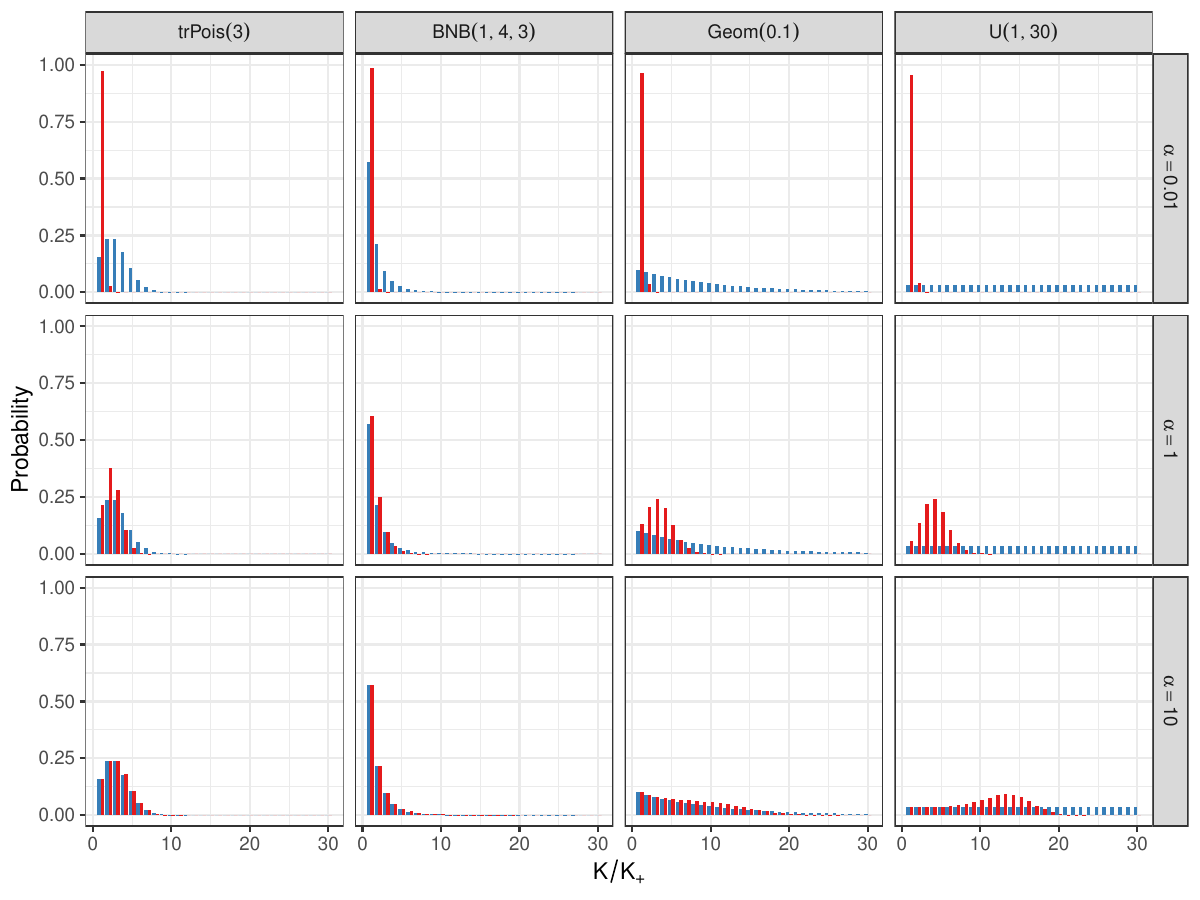}
	\caption{\label{fig:priorKplusDynamic} The prior probabilities of
		$K$ (in blue) and $K_+$ (in red) for the dynamic MFM for different
		priors on $K$ and values for $\alpha$ with $n = 82$.}
\end{figure}

Using $n = 82$ -- the sample size of the Galaxy data set -- the priors
on $K$ (in blue) and on $K_+$ (in red) are visualized by bar plots in
Figure~\ref{fig:priorKplusStatic} for the static MFM and in
Figure~\ref{fig:priorKplusDynamic} for the dynamic MFM. The different
priors on $K$ are in the columns and the values
$\gamma \in \{0.01, 1, 10\}$ and $\alpha \in \{0.01, 1, 10\}$ are in
the rows.  The priors on $K$ are ordered according to the  mean of $K^2$,
i.e., the squared  mean of $K$  plus the  variance of $K$. This results
in 12 combinations of specifications on $(K, \gamma_K)$ in total
inducing different priors on the data clusters $K_+$ for the static as
well as the dynamic MFM. Comparing Figure~\ref{fig:priorKplusStatic}
with Figure~\ref{fig:priorKplusDynamic} indicates that in general the
dynamic MFM leads to priors on $K_+$ inducing stochastically smaller
values.

Figure~\ref{fig:priorKplusStatic} clearly indicates that only
$\gamma = 0.01$ leads to a sparse prior on the number of data clusters
$K_+$ and that the impact of the prior on $K$ increases with
increasing $\gamma$. For $\gamma = 10$, the two priors $p(K)$ and
$p(K_+)$ are essentially the same.
For the dynamic case shown in Figure~\ref{fig:priorKplusDynamic}, the
prior on the number of data clusters $K_+$ induces a very sparse
solution for $\alpha = 0.01$ regardless of the prior on $K$. For
$\alpha = 1$, the prior on $K_+$ is sparser than the prior on $K$ but
the induced prior clearly considerably varies depending on the
selected prior on $K$. For $\alpha = 10$ a close link between the
priors on $K$ and $K_+$ is discernible if the prior on $K$ puts
essentially all mass on small values of $K$, while still considerable
differences between these two priors are visible for the shifted
geometric prior and the uniform prior on $K$ which assign substantial
mass to values $K > 10$.

In summary, if a sparse clustering solution is of interest, also a sparse
prior on $K_+$ should be specified. This can be achieved by specifying
a sparse prior on $K$ and/or small values for $\gamma/\alpha$. In
contrast a flat prior on $K$ (e.g., $\text{U}(1, 30)$) and large
values of $\gamma/\alpha$ will a-priori support large values of $K_+$
(i.e., larger than $4$).

\subsection{Prior parameters $b_0$ and $B_0$ for the component means}
\citet{Richardson+Green:1997} proposed to use empirical Bayes
estimates for $b_0$ and $B_0$ which correspond to the midpoint of the
observed data range for $b_0$ and the squared length of the observed
data range $R^2$ for $B_0$. This choice makes the prior invariant to
the scaling of the data, i.e., invariant to the units of the data used
or standardization of the data.  \citet{Richardson+Green:1997} argue
that this is a sensible weakly informative prior which does not
constrain the component means and does not encourage mixtures with
close component means. They perform a sensitivity analysis for this
prior by considering values ranging from $R^2/10^2$ to $R^2$ for $B_0$,
indicating for the Acidity data set
\citep{Crawford+DeGroot+Kadane:1992} that the estimated number of components are
inverse U-shaped, by first increasing with increasing values for $B_0$
and then decreasing again.

In the following we also use the midpoint of the data for $b_0$.  For
$B_0$ we vary the values to assess the impact on the estimated number
of data clusters by considering the values
$B_0 \in \{6.3, 20, 100, 630\}$. The extreme values correspond to the
limits $R^2/10^2$ and $R^2$ considered by
\citet{Richardson+Green:1997}, 20 corresponds to the empirical
variance of the data and \citet{Phillips+Smith:1996} used 100 in their
analysis.

\begin{figure}[t!]
	\centering
	\includegraphics[width=0.8\textwidth]{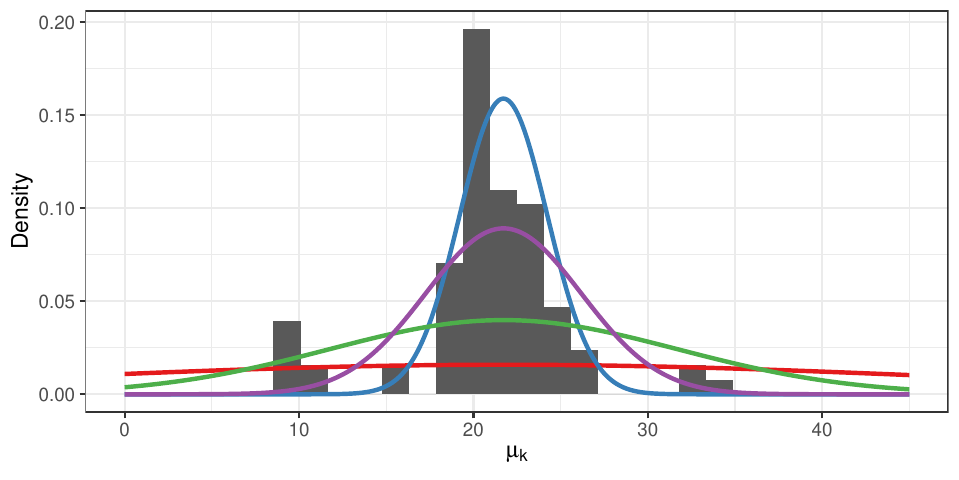}
	\caption{\label{fig:priorMeans} The prior distributions for the
		component means $\mu_k \sim N(b_0, B_0)$ with $b_0$ equal to the
		data midpoint and $B_0 \in \{6.3, 20, 100, 630\}$,
		represented by the blue, purple, green and red line
		respectively, together with a histogram of the Galaxy data
		set.}
\end{figure}

Figure~\ref{fig:priorMeans} visualizes these prior distributions for
the component means together with a histogram of the Galaxy data
set. $B_0 = R^2 = 630$ induces a flat, weakly informative prior as
suggested by \citet{Richardson+Green:1997} with approximately the same
prior density values assigned to all data values observed.
$B_0 = R^2 / 100 = 6.3$ induces the tightest prior for the component
means and assigns very low prior density values to the extreme data
values, thus shrinking the prior component means towards $b_0$. The
smallest value for $B_0$ seems problematic as hardly any weight is
assigned to values below 15 or above 30, where, however, the histogram
would suggest that the centers of small data clusters are located. We
consider this rather extreme range of $B_0$ values to assess
whether the inverse U-shape for the estimated number of data
clusters can also be observed for the Galaxy data set.

\subsection{Prior parameters $c_0$ and $C_0$ for the component variances}

\citet{Richardson+Green:1997} propose to use
$\sigma^{-2}_k \sim \mathcal{G}(c_0, C_0)$ with a hierarchical prior
on $C_0$, but also assess differences in results for a fixed and a
random $C_0$. As we are interested in assessing the impact of
different prior specifications, we only consider the case of fixed
values for $C_0$. Following \citet{Escobar+West:1995},
\citet{Phillips+Smith:1996} and \citet{Richardson+Green:1997}, we use
$c_0 = 2$. We consider $C_0 \in \{0.5, 1, 5, 12.5\}$, where
$C_0 = 0.5$ is used in \citet{Phillips+Smith:1996}, $C_0 = 1$ in
\citet{Escobar+West:1995}, and $C_0 = 12.5$ corresponds to the mean
value considered for the random $C_0$ in
\citet{Richardson+Green:1997}.

\begin{figure}[t!]
	\centering
	\includegraphics[width=0.8\textwidth]{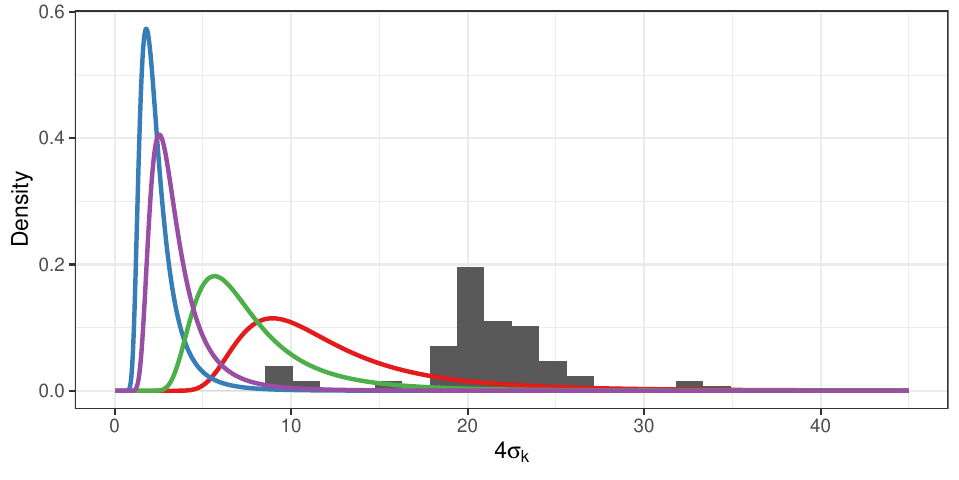}
	\caption{\label{fig:prior4sd} The prior distributions for
		$4 \sigma_k$ induced by the prior on the component precisions
		$\sigma_k^{-2} \sim \mathcal{G}(c_0, C_0)$ with $c_0 = 2$ and
		$C_0 \in \{0.5, 1, 5, 12.5\}$, represented by the blue,
		purple, green and red line respectively, together with a
		histogram of the Galaxy data set.}
\end{figure}

Figure~\ref{fig:prior4sd} visualizes these prior distributions for the
component variances together with a histogram of the Galaxy data
set. The priors induced for $4\sigma_k$ are visualized. These values
correspond to the length of the 95\% prediction interval for a single
component and might be thus seen as representing the volume considered
for the components and hence reflect the prototypical shape imposed
for the clusters. Clearly $C_0 = 0.5$ or $C_0 = 1$ induce prior
standard deviations which allow to include components able to capture
the extreme observations in data clusters of their own, whereas
$C_0 = 12.5$ suggests to approximate the data with overlapping
component distributions. Small values of $C_0$  induce a fine-grained
density approximation, whereas large values of $C_0$ lead to a coarse
density approximation and hence we expect the number of estimated data
clusters to decrease for increasing $C_0$.

\section{Posterior inference}\label{sec:posterior-inference}

In order to obtain samples of the entire 
parameter vector, which
consists of $K$ and, conditional on $K$, of
$\bm{\eta} = (\eta_k)_{k = 1,\ldots,K}$,
$\bm{\mu} = (\mu_k)_{k = 1,\ldots,K}$, and
$\bm{\sigma}^2 = (\sigma_k^2)_{k = 1,\ldots,K}$, from the posterior
distribution, a transdimensional sampler is required which is able to
sample parameter vectors of varying dimension.  We use the telescoping
sampler proposed by
\citet{Fruehwirth-Schnatter+Malsiner-Walli+Gruen:2020}. This MCMC
sampling scheme includes a sampling step where $K$ is explicitly
sampled as an unknown parameter, but otherwise requires only sampling
steps used for finite mixtures.

The posterior inference uses data augmentation and also samples the
component assignments $\bm{S} = (S_i)_{i=1,\ldots,n}$. These latent component
assignments induce random partitions of the data. Thus the sampling scheme also allows to
directly obtain the posterior distribution of the partitions
$\mathcal{C} = \{\mathcal{C}_1, \ldots, \mathcal{C}_{K_+}\}$ of the
data and the induced number of data clusters $K_+$, with
$\mathcal{C}_k$ being the index set of observations assigned to the
$k$th group of the partition $\mathcal{C}$.
To illustrate the connection between the component assignments
$\bm{S}$ and the partitions, assume that $K=3$ and
$\bm{S} = (2, 1, 1, 2, 1, 2, 1, 1, 1, 1)$ for $n = 10$ observations.
Then $K_+ = 2$, since no observations are assigned to the third component, 
and the induced partition is given by
$\mathcal{C} = \{\mathcal{C}_1, \mathcal{C}_2\}$ with
$\mathcal{C}_1 = \{2, 3, 5, 7, 8, 9, 10\}$ and
$\mathcal{C}_2 = \{1, 4, 6\}$.

Following \citet{Fruehwirth-Schnatter+Malsiner-Walli+Gruen:2020}, the
sampling steps of the telescoping sampler consist of:
\begin{enumerate}
	\item Update the partition $\mathcal{C}$ by sampling $\bm{S}$ from
	$p(\bm{S} | \bm{\eta}, \bm{\mu}, \bm{\sigma}^2, \bm{y})$ given by
	\begin{align*}
	P(S_i=k | \bm{\eta}, \bm{\mu}, \bm{\sigma}^2, y_i) \propto \eta_k f_N(y_i|\mu_k,\sigma^2_k).
	\end{align*}
	Determine $N_k=\#\{i=1,\ldots,n|S_i=k\}$ for $k=1, \ldots,K$, i.e.,
	the number of observations assigned to $\mathcal{C}_k$, the $k$th
	group in the partition $\mathcal{C}$ and the number
	$K_+ = \sum_{k=1}^K I\{N_k>0\}$ of non-empty components with
	$I\{\cdot\}$ the indicator function.
	Relabel the components such that the first $K_+$ components are
	non-empty.
	\item Conditional on $\mathcal{C}$, update the parameters of the
	non-empty components for $k=1,\ldots, K_+$:
	\begin{enumerate}[(a)]
		\item Draw the component-specific precisions from the posterior:
		\begin{align*}
		\sigma_k^{-2} | \mu_k, \mathcal{C}, \bm{y} &\sim \mathcal{G}(c_k, C_k),
		\end{align*}
		with
		\begin{align*}
		c_k &= c_0 + \frac{N_k}{2},  
		&C_k &= C_0 + \frac{1}{2} \sum_{i \in \mathcal{C}_k} (y_i - \mu_k)^2.
		\end{align*}
		\item Draw the component-specific means from the posterior:
		\begin{align*}
		\mu_k | \sigma^{-2}_k, \mathcal{C}, \bm{y} & \sim \mathcal{N}(b_k, B_k),
		\end{align*}
		with
		\begin{align*}
		b_k &= B_k (B_0^{-1}b_0 + \sigma_k^{-2} N_k\bar{y}_k),&B_k &= (B_0^{-1} + N_k \sigma_k^{-2})^{-1},
		\end{align*}
		where $\bar{y}_k$ is the sample mean of the observations assigned
		to $\mathcal{C}_k$.
	\end{enumerate}
	\item Conditional on $\mathcal{C}$, draw a new value of $K$ using
	\begin{align*}
	p(K|\mathcal{C}) &\propto p(\mathcal{C}|K)p(K)\propto
	\frac{K!}{(K-K_+)!}
	\frac{\Gamma(K\gamma_K)}{\Gamma(K \gamma_K +N)}	
	\prod_{k=1}^{K_+} \frac{\Gamma(N_k+ \gamma_K)}{\Gamma(1+ \gamma_K)} p(K).
	\end{align*}
	\item Add $K-K_+$ empty components with component-specific parameters
	drawn from the priors:
	\begin{align*}
	\mu_k & \sim \mathcal{N}(b_0, B_0),&\sigma_k^{-2} &\sim \mathcal{G}(c_0, C_0),
	\end{align*}
	for $k=K_+ + 1, \ldots, K$.
	\item Conditional on $\bm{N} = (N_1, \ldots, N_{K_+}, \bm{0}_{K - K_+})$, with
	$\bm{0}_{K - K_+}$ being a $K-K_+$ vector of zeros, draw a new value of
	$\bm{\eta}$:
	\begin{align*}
	\bm{\eta} | \bm{N} &\sim \mathcal{D}_K(\bm{\gamma}),
	\end{align*}
	with $\bm{\gamma} = (\gamma_k)_{k=1,\ldots,K}$ and
	\begin{align*}
	\gamma_k &= \gamma_K + N_k.
	\end{align*}
\end{enumerate}
Inspecting the details of the sampling scheme provides insights into
how the prior specifications influence the conditional posterior
distributions.

The prior specifications of the component-specific parameters
influence Steps~2 and 4. In Step~2, the updates for $c_k$ indicate
that $2 c_0$ might be interpreted as a prior sample size and $C_0/c_0$
corresponds to the variance assumed for these prior observations. The
choice of $c_0 = 2$ thus corresponds to adding 4 observations a-priori
to each component with a variance of $C_0 / 2$. If $C_0 / 2$ is larger
than the empirical within-cluster variance, then $C_k$ is increased
leading to the sampling of inflated $\sigma^2_k$ values. This in turn
induces more overlap across the component densities and thus
potentially leads to a sparser clustering solution with less
data clusters estimated.

The updates for $b_k$ indicate that $b_k$ results as a weighted mean
of the prior value $b_0$ and the mean of the observations currently
assigned to the cluster. According to the formula for $B_k$, the
influence of $B_0$ decreases for data clusters containing many
observations, as the second summand increases with $N_k$. It is also
clear that there is an interaction with the estimate for the
component-specific variance, with larger variances allowing the
component-specific means to vary more in the posterior updates. For
the largest values of $B_0$ considered, we expect that the prior
influence is negligible, and that the posterior updates are only
influenced by the data points currently assigned to this cluster.

Step~3 is influenced by the choice of the prior on $K$ and
$\gamma_K$. More details on this step are given in
\citet{Fruehwirth-Schnatter+Malsiner-Walli+Gruen:2020}. The new $K$ is
sampled from a discrete distribution with support $K \ge K_+$. This
distribution is the more spread out the more the prior on $K$ puts
mass on larger values of $K$ and the smaller $\gamma_K$ is. In
addition the distribution depends on $K_+$ and the cluster sizes
$(N_1,\ldots, N_{K_+})$. This step allows for the birth and death of
empty components.

In Step~4 the parameters of the component-specific distributions of
the new empty components are drawn from the priors.  ``Unattractive''
empty components result in particular when $B_0$ is large and $C_0$ is
small. In this case the sampled $\mu_k$ can be located far away from
the data and the probability that observations are assigned to this
empty component is extremely small in the following Step~1.  Thus, the
``attractiveness'' of the empty components influences whether new
empty components are filled and thus, whether the number of filled
components increases.

Step~5 is influenced by the choice of $\gamma_K$. In particular for
empty components, the value of the Dirichlet parameter only depends on
this prior value, influencing the value $\eta_k$ drawn for these
components and hence also the probability of such an empty component
having observations assigned in Step~1. The smaller $\gamma_k$, the
smaller the sampled $\eta_k$ and thus the smaller the probability that
an observation will be assigned to this component in Step~1.
Furthermore, it can be seen that the prior sample size is equal to
$K\gamma_K$. Thus, for the dynamic MFM where $\gamma_K = \alpha / K$ the prior sample size is
constant over mixtures with different number of components, whereas
for the static MFM where $\gamma_K \equiv \gamma$ the prior sample size linearly increases with the  number of components.

\section{Assessing the impact of different prior
	specifications for the Galaxy data set}\label{sec:assess-impact-diff}

After discussing in detail how the prior specifications might affect
the posterior of the number of data clusters, the following analysis
investigates whether these theoretical considerations can be
empirically verified for the Galaxy data set. The MFM model is
fitted to the Galaxy data set with 384 different prior settings, using
four different specifications of the prior on $K$, using either the
static or the dynamic MFM, considering three different values for the
Dirichlet parameter and four different parameters each for $B_0$ and
$C_0$ in a full factorial design.

\subsection{MCMC estimation}

For each prior setting, posterior inference is performed based on
200,000 iterations after 10,000 burn-in iterations with every fourth
draw being recorded (i.e., a thinning of four). Initially 10
components are filled. The MCMC algorithm is initialized by specifying
values for the component weights and the component-specific
parameters. Equal component weights are specified and all
component-specific variances $\sigma^2_k$, $k=1,\ldots,10$ are set
equal to $C_0/2$. The component-specific means $\mu_k$ are set equal
to the centroids obtained when applying the $k$-means algorithm to
extract 10 clusters from 
the data set. The MCMC iterations start with
Step~1 by assigning observations to the 10 components according to
their a-posteriori probabilities.
%

Partitions are label-invariant. Hence also the number of data clusters
or filled components is a label-invariant quantity and it is not
necessary to resolve the label switching problem
\citep{Redner+Walker:1984} for the following analysis of the results.

\subsection{Analysis of results}

The analysis of the results focuses on the impact of the prior
specifications on the posterior $p(K_+|\bm{y})$ of the number of data
clusters.  The mode of $p(K_+|\bm{y})$ is used as point estimator. In
addition, the entropy of the posterior of $K_+$ is determined to
indicate how informative this posterior is for a point estimate of
$K_+$. The entropy of a discrete random variable $X$ with possible
outcomes $x_1,\ldots, x_I$ is given by
$ - \sum_{i=1}^I P(X = x_i) \log(P(X = x_i) )$. Thus, a high entropy
value for the posterior of $K_+$ indicates rather equal posterior
probabilities for the different values of $K_+$, while a low entropy
value results if the posterior is concentrated on a few values.

The marginal impact of each of the prior specifications on the
estimated number of data clusters $K_+$, based on the posterior mode,
is assessed by averaging the results across all other prior settings.
Table~\ref{tab:marginals} shows the corresponding results. On average,
the estimated number
of data clusters $K_+$ (a) is higher for the static than the dynamic
MFM, (b) increases for increasing values of the Dirichlet parameter,
(c) is lowest for the truncated Poisson prior followed by the BNB(1,
4, 3) prior and, after a substantial gap, followed by the
$\text{Geom}(0.1)$ and finally the uniform $\text{U}(1, 30)$
prior. For the priors on the component-specific parameters, a
non-monotonic influence is indicated for $B_0$.  The average number of
estimated data clusters $K_+$ is highest for $B_0 = 20$, comparable
in-between results are obtained for $B_0 = 6.3$ and $B_0 = 100$, and a
substantial lower average number of data clusters $K_+$ is estimated
for $B_0 = 630$. The influence of $C_0$ on the average number of data
clusters estimated is monotonic and the number substantially decreases
for increasing values of $C_0$. The marginal effects observed in
Table~\ref{tab:marginals} are in line with our prior expectations
based on theoretic considerations and previous results.

\begin{table}[t!]
	\centering
	\begin{tabular}{@{\extracolsep{4pt}}l@{}ll@{}ll@{}ll@{}ll@{}l}
		\hline\\[-8pt]
		{MFM}&$\hat{K}_+$&{$\gamma$ / $\alpha$}&$\hat{K}_+$&{$p(K)$}&$\hat{K}_+$&{$B_0$}&$\hat{K}_+$&{$C_0$}&$\hat{K}_+$\\
		\cline{1-2} \cline{3-4} \cline{5-6} \cline{7-8} \cline{9-10}
		static & 5.89 & 0.01 & 2.98 & trPois(3) & 3.99 & 6.3 & 5.39 & 0.5 & 6.93 \\
		dynamic & 4.70 & 1 & 5.56 & $\text{BNB}(1, 4, 3)$ & 4.35  & 20 & 6.69 & 1 & 6.21 \\
		&  & 10 & 7.33 & Geom(0.1) & 6.00 & 100 & 5.20 & 5 & 4.53 \\
		&  &  &  & U(1, 30) & 6.82 & 630 & 3.90 & 12.5 & 3.50 \\
		\hline
	\end{tabular}
	\caption{Galaxy data set. Average number of estimated data
		clusters $K_+$, based on the mode, marginally for each of the
		different prior specifications.\label{tab:marginals}}
\end{table}

\begin{figure}
	\begin{adjustbox}{addcode={\begin{minipage}{\width}}{
					\caption{Galaxy data set. Estimated number of data clusters
						$K_+$, based on the mode, for different prior specifications. In the rows, the
						results for the static and dynamic MFM are reported, in
						the columns for $\gamma$ or $\alpha \in \{0.01,1,10\}$,
						respectively.\label{fig:mode}}
			\end{minipage}},rotate=90,center}
		\includegraphics[width=\textheight]{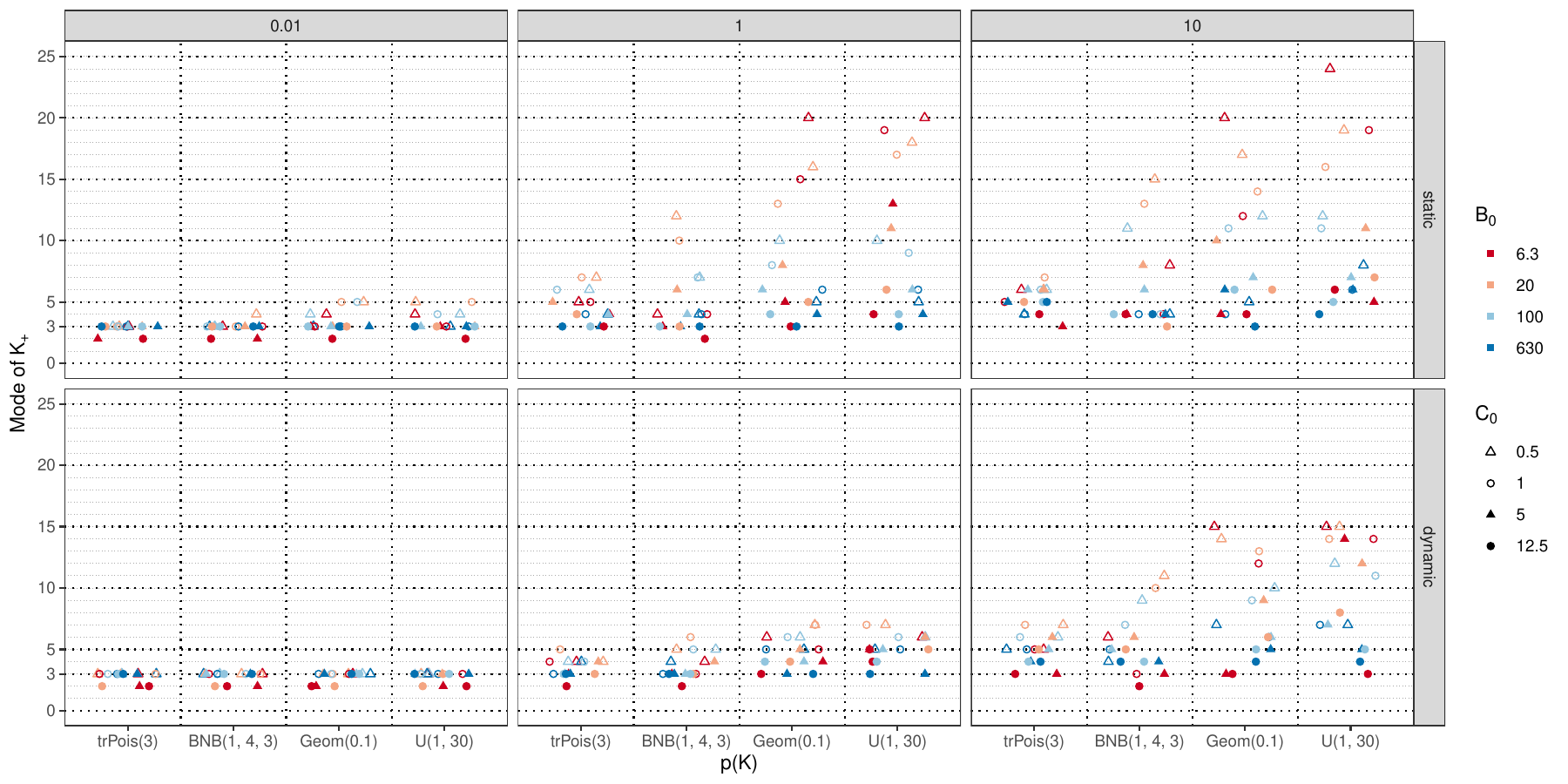}
	\end{adjustbox}
\end{figure}

Figure~\ref{fig:mode} visualizes the results obtained for the 384
different settings in more detail. This figure allows not only to assess
marginal effects, but also to gain insights into the interaction
between the prior specifications.  For each prior setting, the number
of data clusters $K_+$ estimated based on the posterior mode is
indicated by a dot with the value being shown on the $y$-axis. The
results are split into six panels where the top panels contain the
results for the static MFM, while the bottom panels contain the
results for the dynamic MFM. The columns represent the different
values selected for the Dirichlet parameter, $\alpha$ for the dynamic
MFM and $\gamma$ for the static MFM, with values 0.01, 1, and 10 (from
left to right). Within each of the panels the results are grouped on
the $x$-axis by the prior $p(K)$. The priors $p(K)$ are ordered by
their prior mean of $K^2$. Colors and point characters are used to
indicate the different settings used for the component-specific
parameters. Small values of $B_0$ are in red, large values of $B_0$
are in blue. The highly saturated colors indicate the extreme values
of $B_0$ and lighter colors are used for the middle values of
$B_0$. Filled shapes represent the large values of $C_0$, empty shapes
are used for the small values of $C_0$.

Focusing on the dynamic MFM with $\alpha= 0.01$ (in the bottom left
panel), one can clearly see that for nearly all settings the number of
data clusters $K_+$ are estimated to be equal to 3. Only for some
cases, an even smaller number of data clusters $K_+ = 2$ is
estimated. This only occurs for settings where $B_0$ is small and
$C_0$ is large. This suggests that in this panel, where the dynamic
MFM with a sparsity inducing parameter $\alpha$ is fitted, a sparse
clustering solution is obtained regardless of prior on $K$ and also
quite unaffected by the specification on the component-specific
parameters.

The results for the static MFM with $\gamma = 0.01$ are shown above
this panel (in the top left panel). Clearly the sparsity inducing
prior used for $K_+$ leads to the number of data clusters being
estimated as equal to three for most settings. Only for very few
settings, a lower or a higher number of data clusters than 3 (i.e., 2,
4, or 5) is estimated. Again a lower number of data clusters is only
observed in the case where $B_0$ is small and $C_0$ is large. The
higher number of data clusters is observed for small values of $C_0$
and middle values of $B_0$.

Overall the results for $\alpha = 0.01$ for the dynamic MFM and
$\gamma = 0.01$ for the static MFM indicate that the prior on $K$ is
not very influential, as regardless of the choice of the prior on $K$
a sparsity inducing prior for $K_+$ is imposed where a rather large
gap between $K$ and $K_+$ a-priori is likely to occur. Also the
results are quite insensitive to the selection of the parameters for
the component-specific distributions. This implies that if a sparse
clustering solution is desired, one clearly needs to use a small value
for the Dirichlet parameter. The results are rather insensitive to the
specification of the other priors. If the cluster analysis aims at
answering the question what is the minimum number of data clusters
necessary to approximate the data distribution reasonably well, such a
sparsity inducing prior is warranted. In this case the question how
many data clusters are in the Galaxy data set would also be rather
unambiguously answered by three.

Increasing $\alpha$ and $\gamma$ to 1 indicates that the influence of
the other prior specifications on the estimated number of data
clusters increases (middle panels). The dynamic MFM tends to estimate
less data clusters than the static MFM. The difference to
the static MFM becomes more pronounced if the prior on $K$ puts more
mass on the tails.  For the dynamic MFM, all estimated number of data
clusters are at most 7, with higher numbers being more likely for the
uniform and the geometric prior, followed by the BNB prior and the
truncated Poisson prior. Under the static MFM extremely large values
are obtained for the uniform and the geometric prior, with estimates
as large as 20. These large values are obtained if small values are
used in the prior specification for $B_0$ and $C_0$.

For the dynamic MFM, a higher number of data clusters $K_+$ is
estimated for $\alpha = 10$ compared to $\alpha = 1$, while for the
static MFM, rather similar results are obtained for $\gamma = 1$ and
$\gamma = 10$ (panels on the right). For the uniform and geometric
prior on $K$ the estimated number of data clusters varies most,
regardless of whether a static or dynamic MFM is fitted. The prior on $K$
is not particularly sparsity inducing and thus the prior on the
component-specific parameters influences which approximation of the data
density is selected. Small values for $B_0$ induce the most extreme
values for the estimated number of data clusters, with large values of
$C_0$ leading to small numbers and small values of $C_0$ encouraging large
numbers of data clusters.

\begin{figure}
	\begin{adjustbox}{addcode={\begin{minipage}{\width}}{
					\caption{Galaxy data set. Entropy of the posterior of $K_+$
						for different prior specifications. In the rows, the
						results for the static and dynamic MFM are reported, in
						the columns for $\gamma$ or $\alpha \in \{0.01,1,10\}$,
						respectively.\label{fig:entropy}}
			\end{minipage}},rotate=90,center}
		\includegraphics[width=\textheight]{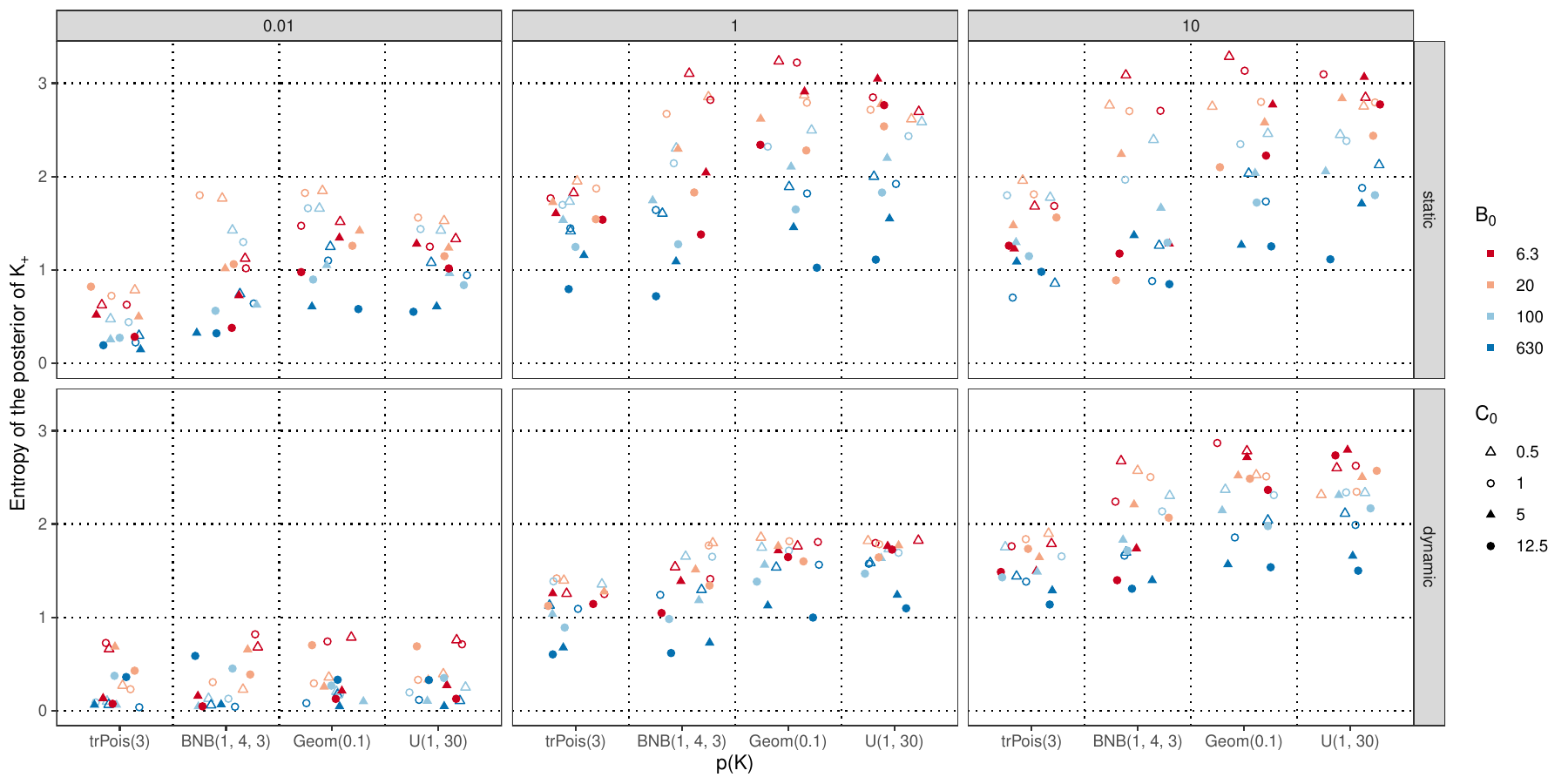}%
	\end{adjustbox}
\end{figure}

Figure~\ref{fig:entropy} visualizes the results obtained for the 384
settings in detail based on the entropy of the posterior of $K_+$. If
the entropy is 0, then all mass is assigned to a single value (which
then also corresponds to the mode shown in Figure~\ref{fig:mode}). For
a fixed support, the uniform distribution has the maximum entropy. For
$\text{U}(1, 30)$, the entropy is $\log(30) \approx 3.40$,
which corresponds to the case where the posterior of $K_+$
assigns the same probability to each value of $K_+$ from one up to
30.

Figure~\ref{fig:entropy} shows that the entropy values are smallest
for the dynamic MFM with $\alpha = 0.01$ with slightly larger values
for the static MFM with $\gamma = 0.01$. For the dynamic MFM, the
entropy increases for increasing $\alpha$. For the static MFM, the
entropy values also increase from $\gamma = 0.01$ to $\gamma = 1$, but
are rather comparable for $\gamma = 1$ and $\gamma = 10$.

Regarding the prior on $K$, smaller entropy values are observed for
the truncated Poisson prior compared to the other priors which have
rather comparable entropy values for a given $\gamma_K$ setting. This
indicates that the smaller prior variance of the prior on $K$ has a
substantial impact on the entropy.

Regarding the prior on $B_0$, a general pattern of the red points
being above the blue points is discernible.  This implies that the
posterior on $K_+$ is particularly spread out for small values of
$B_0$, i.e., where the component-specific mean values are shrunken
towards the midpoint. We conjecture that in this setting posterior
mass is also assigned to small values of $K_+$ as due to the shrinkage
there is posterior support for solutions with few data clusters. For
example, the observations in the Galaxy data set with large values
which seem to form a small data cluster of their own, might be merged
with observations from the middle bulk of the observations due to
shrinkage, inducing a large component-specific variance and thus a
coarse density approximation.

Regarding $C_0$, the general pattern is that the filled shapes are
below the empty shapes, indicating that the entropy increases with
decreasing values of $C_0$.  This means that 
the probability mass is more spread
out if one aims at a fine-grained approximation using a rather small
volume as prototypical shape for the clusters. In particular, if the
aim is semi-parametric density estimation and a small volume is
imposed, it is not to be expected that a single mixture with a 
specific  value of $K_+$ approximates the data distribution well, but
rather a range of mixtures with different values of $K_+$ perform well
and \final{all well fitting mixtures may be combined to obtain a good approximation.}

\section{Assessing the impact of different prior specifications for
	artificial data}\label{sec:assess-gener-artif}

To complement the results obtained for the Galaxy data set, a
simulation study with artificial data is performed where the data
generating process and the true number of data clusters are
known. Results are obtained and compared for maximum likelihood
estimation as well as Bayesian inference with different prior
specifications. In the simulation study also the impact of different
sample sizes and of fitting a misspecified mixture model is assessed.

\subsection{Data generation and analysis setup}

We designed the data generating process in the simulation study to
induce data sets which are similar to the Galaxy data set. The
underlying data generating process is either a mixture of univariate
Gaussian distributions or a mixture of univariate uniform
distributions with four components each.  Two different sample sizes
with $n = 100$ and 1000 data points are considered. For $n = 100$, the
four cluster sizes are fixed to 5, 55, 30 and 10 and these cluster
sizes are multiplied by 10 for $n = 1000$.  For the Gaussian mixture,
the four component means and standard deviations are given by
$\mu_k \in \{9.5, 20, 24.5, 33\}$ and
$\sigma_k \in \{0.25, 1, 1, 0.5\}$, respectively. For the uniform
mixture, the lower and upper bounds of the four uniform component
distributions are given by
$\{(9, 10), (18, 22), (22, 27), (32, 34)\}$.  \final{100 different
	artificial data sets are drawn and analyzed for each of the
	scenarios.}


Results for maximum likelihood estimation are obtained using
the R package \pkg{mclust}. The default initialization scheme
implemented in the package is used and model choice with regard to $K$
is performed using the BIC. Model choice consists
in selecting  the best model within three modeling approaches for the component variances:
(1) equal variances across components, (2) unequal variances across
components, (3) the best model according to the BIC among the equal
and unequal variance models.

The hierarchical MFM model (as given in~\eqref{eq:MFM}) is
fitted to each of the \final{100 artificial data sets of each scenario} using
essentially the same prior specifications as used for the analysis
of the Galaxy data set. We only make two modifications. We restrict
the prior specifications to the extreme values for$B_0$ and
$C_0$, \final{i.e., $B_0 \in \{6.3, 630\}$ and $C_0 \in \{0.5, 12.5\}$,}  to
obtain a more succinct presentation of the results.  Furthermore, a
uniform prior $\text{U}(0, 100)$ for $K$ instead of a uniform prior
$\text{U}(0, 30)$ is specified. Given that larger sample sizes are
considered, a larger upper bound for the uniform distribution is
selected to ensure that the specific bound selected is still
inconsequential.  We base the posterior inference for each
prior setting on MCMC sampling using 200,000 iterations after
discarding 10,000 iterations as burn-in samples and using a thinning
of four. The same initialization scheme as for the Galaxy data set
is employed.

\subsection{Analysis of results}

First, we inspect the results obtained using maximum likelihood
estimation with the BIC for the three modeling approaches for the
different sample sizes and data generating processes. It
should be noted that BIC selects the number of components $K$ rather
than the number of data clusters $K_+$.  The estimated number of
components are summarized in Table~\ref{tab:artificial-mle} for each
setting using the minimum, the 25\%, 50\% and 75\% quantile and the
maximum to characterize the distribution of these estimates across the
100 data sets.

\begin{table}[t!]
	\centering
	\begin{tabular}{@{}l@{\quad}l@{\quad}lll@{}}
		\hline
		& $n$ & Equal & Unequal & Equal or unequal \\
		\hline
		Gaussian &  100 & [4.0, 4.0, 5.0, 5.0, 7.0] & [4.0, 4.0, 4.0, 4.0, 5.0] & [4.0, 4.0, 4.0, 4.0, 5.0]\\
		& 1000 & [6.0, 7.0, 9.0, 9.0, 12.0] & [4.0, 4.0, 4.0, 4.0, 4.0] & [4.0, 4.0, 4.0, 4.0, 4.0]\\
		\hline
		Uniform &  100 & [4.0, 5.0, 5.0, 6.0,  8.0] & [3.0, 4.0, 4.0, 5.0, 7.0] & [3.0, 4.0, 5.0, 5.0, 8.0]\\
		& 1000 & [7.0,  8.8,  9.0,  9.0, 15.0] & [5.0, 6.0, 7.0, 7.0, 9.0] & [5.0, 6.0, 7.0, 7.0, 9.0]\\
		\hline
	\end{tabular}
	\caption{Artificial data, maximum likelihood estimation with the
		BIC.  Results are shown for the three different modeling
		approaches consisting of equal, unequal and equal as well as
		unequal variances for the component distributions. The estimated
		number of components are summarized over 100 data sets by the
		minimum, the 25\%, 50\% and 75\% quantile and the maximum in
		square brackets.\label{tab:artificial-mle}}
\end{table}

If the data are drawn from a Gaussian mixture and the larger sample size  $n=1000$ is considered, maximum likelihood
estimation in combination with BIC always selects four components in case the unequal
variance model is specified or the best model among the equal and
unequal variance models is selected. Only slightly worse results are obtained
for the smaller sample size, $n = 100$, when these modeling approaches
are considered. If the equal variance model is enforced, the number
of components are correctly selected or slightly overestimated for
$n = 100$. For the larger sample size, considering only the equal
variance model leads to overestimating the number of components by
at least two with a median number of five and up to eight
components in addition.

If the mixture model is misspecified, the performance of the maximum
likelihood estimation deteriorates. This is expected as the BIC
takes goodness-of-fit of the estimated density into account to
select a suitable number of components for the mixture
distribution. For the smaller sample size, $n = 100$, the number of
components are only slightly overestimated regardless of the modeling
approach. The estimated number of components increases for the
larger sample size, $n = 1000$. In this case, the correct number of
components is never selected and there are either at least five or
seven components included in the final mixture distribution. The
maximum likelihood estimation approach thus performs poorly if the
model is misspecified and the sample size is rather large.

\begin{table}[t!]
	\centering
	\begin{tabular}{@{\extracolsep{4pt}}l@{}ll@{}ll@{}ll@{}ll@{}ll@{}l}
		\hline
		\multicolumn{12}{c}{Gaussian}\\
		\hline\\[-8pt]
		$n$&$\hat{K}_+$& {MFM}&$\hat{K}_+$&{$\gamma$ / $\alpha$}&$\hat{K}_+$&{$p(K)$}&$\hat{K}_+$&{$B_0$}&$\hat{K}_+$&{$C_0$}&$\hat{K}_+$\\
		\cline{1-2} \cline{3-4} \cline{5-6} \cline{7-8} \cline{9-10}\cline{11-12}
		100 & 5.12 &static & 6.15 & 0.01 & 3.97 & trPois(3) & 4.69 & 6.3 & 6.34 & 0.5 & 6.65 \\
		1000& 5.87 & dynamic & 4.83 & 1 & 5.25 & $\text{BNB}(1, 4, 3)$ & 4.97  & 630 & 4.65 & 12.5 & 4.34 \\
		&  & & &10 & 7.26 & Geom(0.1) & 5.66 &  &  &  &  \\
		&  & & & &  & U(1, 100) & 6.66 &  &  &  &  \\
		\hline
		\multicolumn{12}{c}{Uniform}\\
		\hline\\[-8pt]
		$n$&$\hat{K}_+$& {MFM}&$\hat{K}_+$&{$\gamma$ / $\alpha$}&$\hat{K}_+$&{$p(K)$}&$\hat{K}_+$&{$B_0$}&$\hat{K}_+$&{$C_0$}&$\hat{K}_+$\\
		\cline{1-2} \cline{3-4} \cline{5-6} \cline{7-8} \cline{9-10}\cline{11-12}
		100 & 5.56 &static & 7.33 & 0.01 & 4.85 & trPois(3) & 5.57 & 6.3 & 7.65 & 0.5 & 8.61 \\
		1000& 7.76 & dynamic & 5.99 & 1 & 6.85 & $\text{BNB}(1, 4, 3)$ & 6.27  & 630 & 5.67 & 12.5 & 4.71 \\
		& & & &10 & 8.27 & Geom(0.1) & 6.84 &  &  &  &  \\
		&  & & & &  & U(1, 100) & 7.96 &  &  &  &  \\
		\hline
	\end{tabular}
	\caption{Artificial data, Bayesian estimation. Average number of
		estimated data clusters $K_+$, based on the mode, marginally for
		each of the different prior specifications and whether the component
		distributions are Gaussian or uniform
		distributions.\label{tab:art-marginals}}
\end{table}

Table~\ref{tab:art-marginals} summarizes the results for the
Bayesian approach with different prior specifications when using the same
artificial data as used for maximum likelihood estimation. 
Here, we report inference regarding the number of data clusters $K_+$ rather
than $K$.   The table shows the
marginal effects of the different prior specifications on
the estimated number of data clusters. The effects are again in line with our expectations
and confirm the insights gained for the Galaxy data set. More specifically,
it
can be observed that the number of estimated data clusters increases
for increasing prior mean of $K$ and smaller values of $C_0$. In
addition it can be noted that the model misspecification leads on
average to more data clusters being estimated.

\begin{figure}
	\vspace{-1.7cm}
	\includegraphics[width=\textwidth]{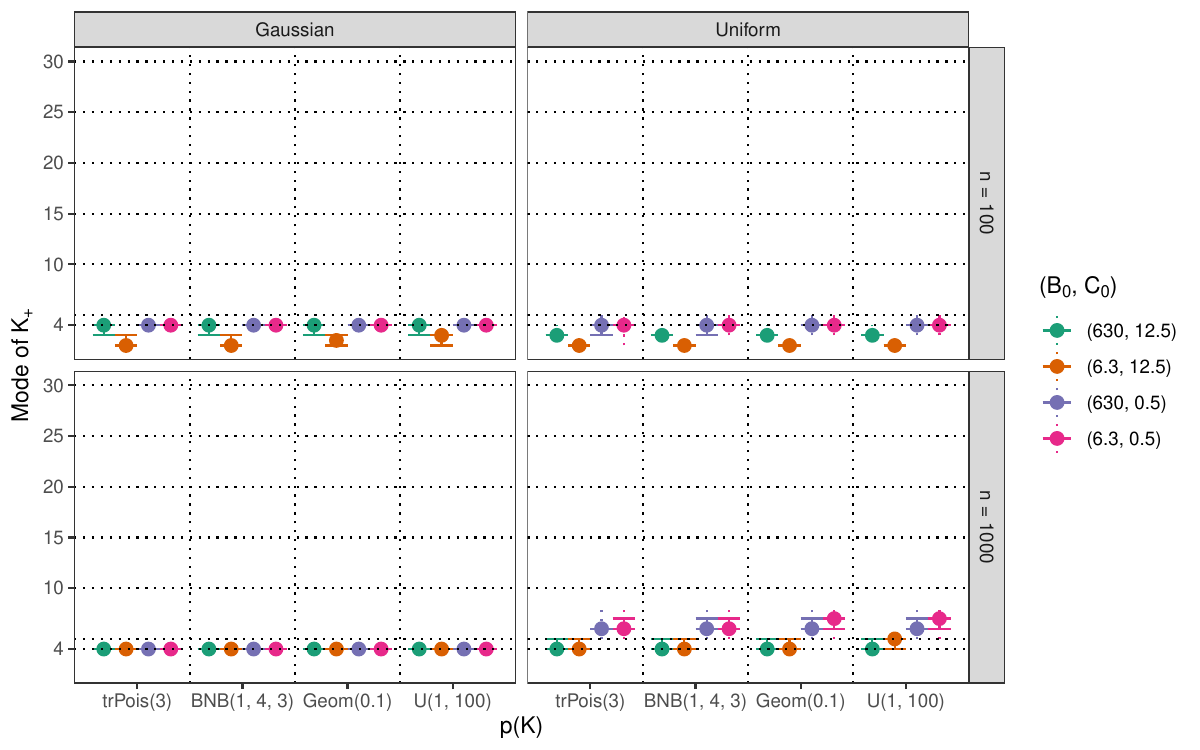}
	\caption{Artificial data, dynamic MFM with $\alpha = 0.01$.
		Estimated number of data clusters $K_+$
		based on the mode for 100
		data sets with different prior specifications for the prior on
		$K$, $B_0$ and $C_0$. In the rows, the results for different
		samples sizes ($n = 100$ or $1000$) are reported, in the columns
		for different data generating processes, mixtures of Gaussians or
		mixtures of uniform distributions.  The results for the
		$(B_0, C_0)$ specifications as listed in the legend are shown from
		left to right within each prior on $K$ setting.
		\label{fig:artificial-results-dynamic}}
	\includegraphics[width=\textwidth]{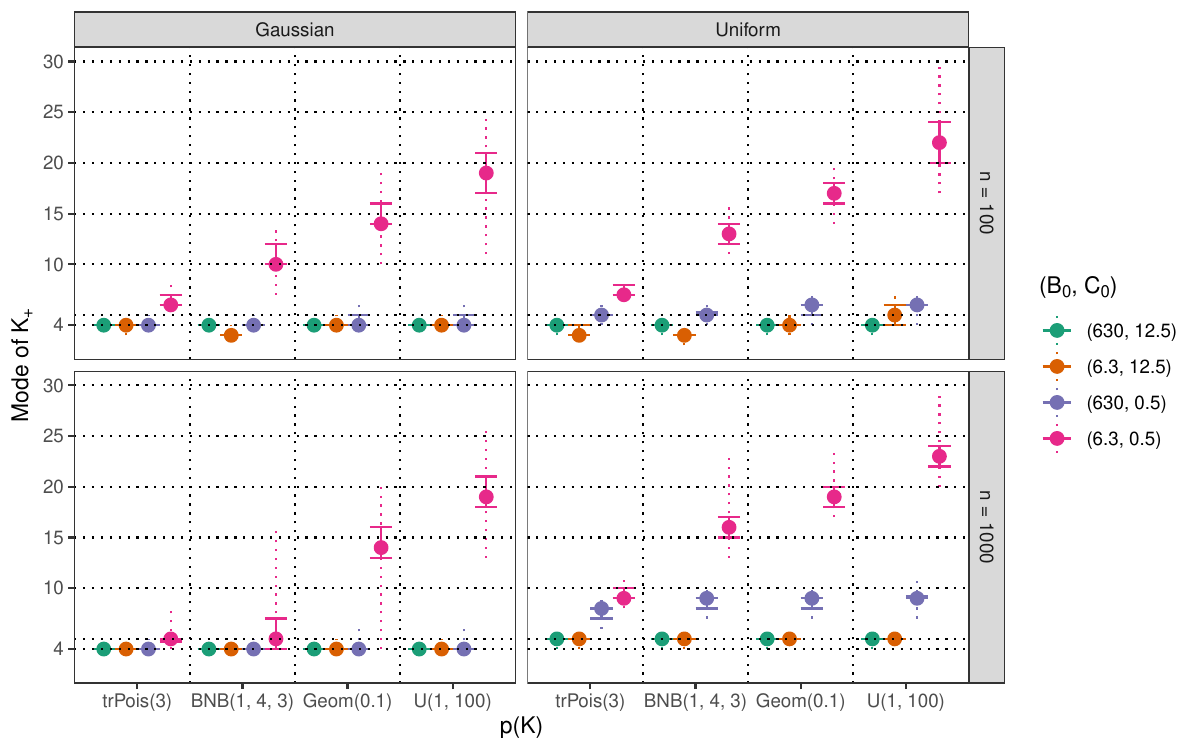}
	\caption{Artificial data, static MFM with $\gamma = 1$.
		Estimated number of data clusters $K_+$ based on the mode for 100 data sets with different prior specifications
		for the prior on $K$, $B_0$ and $C_0$. In the rows, the
		results for different samples sizes ($n = 100$ or $1000$) are reported,
		in the columns for different data generating processes,
		mixtures of Gaussians or mixtures of uniform distributions.
		The results for the $(B_0, C_0)$ specifications as listed in the legend are
		shown from left to right within each prior on $K$ setting.
		\label{fig:artificial-results-static}}
\end{figure}

In the following the impact of the prior settings
on the estimated number of data clusters is investigated in more
detail for a dynamic MFM with $\alpha = 0.01$ and a static MFM with
$\gamma = 1$. In Figures~\ref{fig:artificial-results-dynamic} and
\ref{fig:artificial-results-static}, the results over the 100 data
sets are summarized. The median estimated number of
data clusters $K_+$ is represented by the bullet points. In addition error bars
connected by straight lines indicate the range between the 25\% and
the 75\% quantile, whereas dotted lines indicate the total range from
minimum to maximum. 

Results for the dynamic MFM with $\alpha = 0.01$ shown in
Figure~\ref{fig:artificial-results-dynamic} indicate that if the
data generating process is a Gaussian mixture, the correct
number of data clusters is selected most of the times, in particular
if the sample size is large, i.e., for $n = 1000$.
For the smaller
sample size, $n = 100$, the number of data clusters is
underestimated if large variances are a-priori assumed for the
component distributions and in particular if also the
component-specific means are shrunken together because of the small value
of $B_0$.
Further, it can also be seen that for the larger sample size the estimated number of data
clusters coincide with the true number of data clusters regardless
of the prior distributions used for $K$ and the other parameters. Thus, if there is no model misspecification and the data
set is sufficiently large, using a dynamic MFM with a small $\alpha$
value leads to correct estimates of the number
of data clusters regardless of the other prior settings.

If the component distribution is misspecified but the data
set is small, the results obtained are rather similar to the
Gaussian case. However, for the
larger sample size, $n=1000$, four clusters are only estimated if the priors on the component distributions assume large values
for $C_0$ and $B_0$.  Otherwise, the number of clusters is clearly overestimated
with median values between six and seven. Thus, using a
dynamic MFM with a small $\alpha$ value in combination with sensible
priors on the component distributions results in obtaining the
correct estimates for the number of data clusters even if  the model is misspecified and the data set is rather  large.

The dynamic MFM with a small $\alpha$ value is clearly a
successful strategy for obtaining an estimate of the number of data
clusters which could be seen as the ``minimum number of data
clusters'' being present in the data. To further emphasize the
advantages of this approach, the results for the static MFM with
$\gamma = 1$ are, in comparison, inspected in  
Figure~\ref{fig:artificial-results-static}. Regardless of the data
generating process and the sample size, using priors on the
component distributions which induce small values for $B_0$ and
$C_0$ leads to overestimating the number of data clusters in the
data set. The amount of this overestimation strongly depends on the
prior used for $K$. The estimated number of data clusters in fact
increases with the prior mean of $K$, e.g., for
$K \sim \text{U}(1,100)$ the median number of estimated data
clusters is about 20 regardless of the data generating process and
the sample size.

For Gaussian mixtures, using a static MFM with $\gamma=1$ again leads to correct estimates of the number of data clusters
for almost all prior specifications. The only
exception is the already highlighted setting where the priors on the
component distributions induce small values for $B_0$ and $C_0$.

In contrast, for $n = 1000$ the number of data clusters is always
overestimated in case of model misspecification and a static MFM with
$\gamma = 1$ is fitted.  For large values of $C_0$, consistently
five data clusters are estimated instead of four. Using a small
value for $C_0$ allows for semi-parametric density estimation and
hence leads to a substantial overestimation of the number of data
clusters. Thus, it is not
recommended to use a static MFM with $\gamma=1$ in applications
where the component distributions are likely to misspecify the
cluster distribution and a sparse clustering solution is of
interest.

\section{Discussion and conclusions}\label{sec:conclusions}

In this paper, we respond to the call for action made by
\citet{Aitkin:2001} regarding the need to provide more insights into
the influence of different prior specifications when fitting Bayesian
mixture models. Based on recent developments in the context of MFMs,
we use the model specification of a MFM, considering the static as
well as the dynamic case.
The Galaxy data set is used to illustrate the prior impact on the
estimated number of data clusters $K_+$ using the mode as well as on
the entropy of the posterior of $K_+$. Results confirm the marginal
effects postulated, but also interesting interaction effects are
discerned.

Aiming at a sparse clustering solution using a dynamic MFM
with $\alpha = 0.01$ gives stable results regardless of the prior on
$K$. The clustering solution is also rather insensitive to the prior
on the component-specific parameters as long as they are sensible.
Such a prior is especially recommended to be combined with large
component variances and large variances of the component means, if
the data set is large and the cluster density is unknown and likely
to be misspecified (which is often the case in applications). Such a
prior specification will avoid overfitting and lead to an estimate
of $K_+$ that could be interpreted as the ``minimum number of data
clusters'' being present in the data and in general might provide a
better clustering performance than the maximum likelihood approach
combined with the BIC.

For the Galaxy data set, a dynamic MFM with $\alpha = 0.01$ would lead
to an unambiguous estimate of three data clusters with also the
posterior distributions being rather concentrated on very few
values. This is in line with the conclusion drawn in
\citet{Aitkin:2001} for the maximum likelihood framework using equal
variance components in the mixture model.

We suggest to use the dynamic MFM with small $\alpha$ value and
reasonable component-specific distributions in a Bayesian model-based
clustering application where a minimum number of data clusters is to
be identified. For the component-specific distributions, shrinking the
prior mean is not recommended, whereas for the component-specific
variances using reasonable values is important to guard against too
fine-grained or too coarse approximations. In the univariate case the
visualization of the induced volume (see Figure~\ref{fig:prior4sd}) is
useful to determine a suitable value for $C_0$. A generalization of
such a visual tool to the multivariate case or other
component-specific distributions would be of interest. Further
analysis is also required to gain insights of the prior impact on
Bayesian cluster analysis results for data sets with many variables
and with other component-specific distributions. In addition, if less
focus is given to the clustering aspect of the MFM model, it might
also be interesting to investigate the posterior of the number of
components $K$, in particular based on a simulation study where $K$
and $K_+$ are known and may be manipulated to be different.

\bigskip
\smallskip
\noindent \textbf{Acknowledgements.}
	The authors gratefully acknowledge support from the {Austrian
		Science Fund (FWF)}: P28740, and through the WU Projects grant
	scheme: IA-27001574.

\bibliographystyle{spbasic}   
\bibliography{galaxy}

\end{document}